\documentclass[aps,pre,twocolumn,groupedaddress,longbibliography]{revtex4-1}

\newcommand{\Pe}{\text{Pe}}

\usepackage{graphicx} 
\usepackage{color}
\usepackage{times,mathptmx}
\usepackage{graphicx} 
\usepackage[capitalize]{cleveref}
\begin{document}


\title{Drying dynamics of a charged colloidal dispersion\\ in a confined drop}


\author{Charles Loussert}
\author{Anne Bouchaudy}
\author{Jean-Baptiste Salmon}
\email[]{Jean-Baptiste.Salmon-exterieur@solvay.com}
\affiliation{CNRS, Solvay, LOF, UMR 5258, Univ. Bordeaux, F-33600 Pessac, France}


\date{\today}

\begin{abstract}
We performed a thorough investigation of the drying dynamics of a charged colloidal dispersion drop in a confined geometry. We developed
an original methodology based  on Raman micro-spectroscopy to measure spatially-resolved colloids concentration profiles during the drying of the drop.
These measurements lead to estimates of the collective diffusion coefficient of the dispersion over a wide range of concentration.  
The collective diffusion coefficient is one order of magnitude higher than the Stokes-Einstein estimate showing the importance of the electrostatic
interactions for the relaxation of concentration gradients. At the same time, we also performed fluorescence imaging of tracers embedded within the 
dispersion during the drying of the drop, which reveals two distinct regimes. At early stages, 
concentration gradients along the drop lead to buoyancy-induced flows. Strikingly, these flows
do not influence the colloidal concentration gradients that generate them, as the mass transport remains dominated by diffusion. At longer time scales, the tracers trajectories reveal the formation of a gel which dries quasi homogeneously. For such a gel, we show using linear poro-elastic modeling, that the drying dynamics is still described by  the same transport equations as for the liquid dispersion. However, the collective diffusion coefficient follows a modified generalized Stokes-Einstein relation, as also demonstrated in the context of unidirectional consolidation  by Style {\it et al.} 
  [Crust formation in drying colloidal suspensions, Style {\it et al.}, Proc. R. Soc. A {\bf 467}, 174 (2011)].\end{abstract}


\maketitle

\section{Introduction}

Understanding the drying dynamics of a colloidal dispersion is probably one of the most challenging issue in the field of coating engineering~\cite{Russel:11,Routh:13}.
This intricate process couples both fundamental aspects of colloidal physics and transport phenomena (interactions, stability, convection/diffusion, etc.)~\cite{Goehring:10,Li:2012,Kim:13,Dufresne:03,Bodiguel:10A,Ziane:15a,Lidon:14,Piroird:16}, up to the formation of solids with 
internal mechanical stresses which are released through (often detrimental) instabilities such as shear bands~\cite{Boulogne:14,Goehring2016,Yang:2015,Kiatkirakajorn:15}, film delamination~\cite{Sarkar:11,Xu:13}, and cracks~\cite{Allain:95,Gauthier:07,Giorgiutti-Dauphine2014}. 

The common description of drying in many  geometries (suspended or confined drops, films, etc.)  
is the following: solvent evaporation (at a rate $\dot{E}$) induces convection towards the evaporating air/dispersion interface, thus concentrating the colloids up to the formation of a close-packed colloidal material, see e.g. Refs.~\cite{Routh:04,Boulogne:13,Lintingre,Tsapis:05}. 
This concentration process is mainly governed by the competition between drying-induced convection in the bulk (at a velocity $\dot{E}$ relatively to the evaporating interface) and collective diffusion relaxing concentration gradients. The scale of the expected gradient  $\xi \sim D_0/\dot{E}$ where $D_0$ is the  colloid diffusion coefficient, is often 
compared to a characteristic length scale $L$ (e.g. film thickness) using the P\'eclet number $\Pe = \dot{E} L / D_0$. In most experiments,  Stokes-Einstein estimates for $D_0$ point out that the drying of dispersions leads to the formation of a thin crust at the evaporating interface, e.g. $\xi \simeq 10$--100~$\mu$m for particles with radii $a = 10$--100~nm dispersed in water, and for $\dot{E} \simeq 100~$nm/s.

However, highly charged colloids repeatedly used as model systems in this wide research area (e.g. the commercial Ludox silica nanoparticles in some of the above cited works), deserve further attention.
Indeed, long range electrostatic interactions (as compared to the colloid radii $a\sim 10$~nm) may lead to a collective diffusion coefficient 
much larger than the Stokes-Einstein prediction $D_0$. 
The collective diffusion coefficient $D(\varphi) $ indeed results from an interplay between colloidal interactions 
and hydrodynamic interactions at finite concentrations~\cite{Russel,Nagele}, and it follows the generalized Stokes-Einstein relation Eq.~(\ref{eq:GSE}), see later Sec.~\ref{sec:discussion} for a more detailed discussion. 
Higher values than $D_0$ can thus be observed in colloidal dispersions with high osmotic compressibility (strong repulsive interactions) and low hydrodynamic friction of the relative flow solvent/particles.
Such phenomena are well-known by the community investigating the dynamics of colloids, and it has been reported many times using scattering techniques, see e.g. Refs.~\cite{Walrand86,Petsev92,Gapinski:07}. Other  evidences were even reported using measurements of concentration gradients in the context of unidirectional drying using in-situ SAXS~\cite{Boulogne:14,Goehring2016}, or by measurements of permeate flux in ultra-filtration experiments~\cite{ROA2016}. Moreover,  a transition from a liquid to a poro-elastic gel often occurs for these highly charged colloids at concentrations well below the close-packing~\cite{DIGIUSEPPE2012,Bodiguel:10A,Boulogne:14,Ziane:15a}. Such a transition was shown to play a crucial role for the formation of shear bands in unidirectional drying~\cite{Boulogne:14,Yang:2015,Kiatkirakajorn:15}.  Despite the numerous works employing these charged nanoparticles to investigate the drying of dispersions, a {\it quantitative} description of the concentration process taking into account  these phenomena is still missing. 

In the present work, we provide spatially-resolved measurements of the colloid concentration during the drying  of  a charged nanoparticles dispersion in a model geometry: a confined drop between two circular plates, see Fig.~\ref{fig:setup}. These original measurements based on Raman micro-spectroscopy enable us to evidence that evaporation induces slight concentration gradients within the drop owing to enhanced collective diffusion. Moreover, these spatially-resolved measurements of the concentration fields make it possible to extract precise estimates of the collective diffusion coefficient $D$ over the colloid volume fraction range $\varphi \approx 0.25$--0.6. The measured values of $D(\varphi)$ are about one order of magnitude higher than $D_0$, the Stokes-Einstein estimate. We also combine these measurements with fluorescence imaging of tracers  within the dispersion. These experiments reveal 
two distinct regimes before the consolidation. At low colloid volume fraction $\varphi \leq 0.3$, 
the concentration gradients lead to buoyancy-induced flows as the density of the dispersion  evolves with $\varphi$. Strikingly, we show that these flows do not influence the concentration gradients that generate them, as colloidal transport remains dominated by diffusion in such confined geometries~\cite{Selva:12}.
For $\varphi \geq 0.3$, these buoyancy-induced flows vanish, and the trajectories of the tracers suggest the formation of a gel which dries quasi homogeneously. For such a poro-elastic gel, it has recently been shown  by Style {\it et al.} that the collective diffusion coefficient does not strictly follow  the generalized Stokes-Einstein relation as in a liquid dispersion~\cite{Style:11}. This result, due to the constraints imposed by the geometry, was demonstrated theoretically in the context of unidirectional consolidation using non-linear poro-elasticity~\cite{MacMinn2016}. We find again the same result in our experimental configuration but using linear poro-elastic modeling, as solvent evaporation induces only slight concentration gradients during the drying of the drop.

\section{Confined drying \label{sec:confineddrying}}

\begin{figure}[htbp]
\begin{center}
\includegraphics{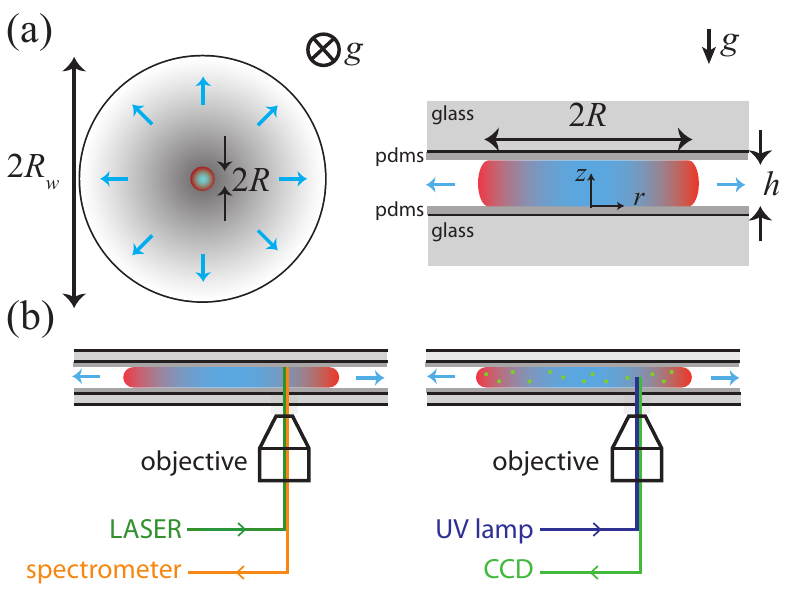}
\caption{(color online) (a) Schematic (top and side) views of the experiments: a drop of an aqueous colloidal dispersion is confined between two 
circular wafers separated by a fixed height $h$. 
A thin PDMS layer coated on each inner side of the glass wafers prevents from the pinning of the receding contact line.
Colors code the drying-induced concentration gradient of colloids, $\mathbf{g}$ indicates the gravity direction, and blue arrows show the diffusive drying  within the cell. 
(b, left) Raman imaging: a laser is focused at a given location within the drop, and the 
scattered light is collected using a spectrometer. The drop is then scanned along its  diameter using a motorized stage.
(b, right) Fluorescence microscopy: fluorescent tracers (green symbols) are dispersed within the drop
to visualize flows.}
\label{fig:setup}
\end{center}
\end{figure}

Figure~\ref{fig:setup}(a-b) illustrates the geometry of confined drying: a  drop is squeezed between two circular wafers separated by small spacers of fixed height $h$~\cite{Clement:04}. 
In the following experiments, the typical drop volume ranges from $\simeq  0.5$ to 2.5$~\mu$L, the wafer radius $R_w = 3.81$~cm, $h$ ranges from 80 to 250~$\mu$m, and the initial drop radius is $R_0 \simeq 1$~mm.   
This model geometry offers numerous advantages as compared to free films or sessile droplets: 
(i) drying rates are mainly governed by a diffusive mass transfer and thus by geometrical parameters, (ii)
axis-symmetry combined with confinement provides both simplified experimental observations and easy modeling, and (iii) the limited free surface hinders Marangoni instabilities.
This geometry recently emerges  for these reasons as a powerful technique for investigating quantitatively the drying of complex fluids: formation of crusts in hard-sphere dispersions~\cite{Leng:10} and the associated {\it buckling} instability~\cite{Pauchard:11,Boulogne:13}, drying of anisotropic colloidal dispersions~\cite{Yunker:12}, and even dynamic investigation of phase diagrams of copolymer solutions~\cite{Daubersies:12}.
A complete theoretical description of the drying kinetics (including the case of pure solvents and binary mixtures) can be found in Refs.~\cite{Daubersies:11,Daubersies:12}, and we only summarize below the information needed in the following. 

The drying kinetics of a pure water drop is limited in this geometry by the quasi-static diffusion of the vapor within the cell. The evaporative flux at the air/drop interface is $J(R) = -D_g c_s(1-a_e)/ [R\ln(R/R_w)]$, where $c_s$ is the concentration of saturated water vapor pressure in air
($c_s \simeq 1~$mol/m$^3$), $a_e$ the external residual humidity,  and $D_g$ the diffusion coefficient of the vapor in the gas phase. Mass conservation  leads to the following temporal evolution for the drop radius: 
\begin{equation}
\frac{\text{d} R}{\text{d}t} = \frac{\tilde D}{R \ln{(R/R_w)}}, 
\label{eq:sechage}
\end{equation}
with the effective diffusion coefficient $\tilde D = D_g \nu_s c_s(1-a_e)$ and $\nu_s$ the molar
volume of water ($\nu_s \approx 1.8\times 10^{-5}$~m$^3$/mol). 
For the room conditions investigated in the present work ($T=18$--21$^\circ$C, $a_e=0.4$--0.65),  $\tilde D$ ranges from 2.2 to $3\times 10^{-10}$~m$^2$/s
and the velocity of the receding meniscus is of the order of  $\sim \tilde D/ (R \ln(R/R_w)) \simeq 100$~nm/s for radii $R \simeq 1$~mm .
Note that for small drops radius $R \ll R_w$, the exact geometry imparted by the cell only plays a logarithmic role 
through the term $\ln{(R/R_w)}$ (i.e. increasing the wafer's size only decreases logarithmically the drying kinetics).

We further define the normalized area $\alpha(t) = [R(t)/R_0]^2$, with $R_0 = R(t=0)$. The solution of Eq.~(\ref{eq:sechage}) is:
\begin{eqnarray}
4 \tilde D t / R_w^2= \beta\alpha\left[\ln(\beta\alpha)-1\right]-\beta\left[\ln(\beta)-1\right]\,, \label{eq:tauf}
\end{eqnarray}  
with $\beta = [R_0/R_w]^2$. 

We now turn to the case of colloidal dispersions, and we  briefly summarize the 
classical description of mass transport within such binary mixtures. We consider (incompressible) colloids of radii $a$ dispersed in water, and we define $\varphi$ their volume fraction. 
We then define $\mathbf{v_f}$ and $\mathbf{v_s}$  the average velocities  of the solvent and colloids respectively. 
Conservation equations are:
\begin{eqnarray}
&&\partial_t \varphi + \nabla. (\varphi \mathbf{v_s}) = 0\,,\label{eq:masscons1}\\ 
&&\partial_t (1-\varphi) + \nabla. (1-\varphi) \mathbf{v_f} = 0\,, \label{eq:masscons2}
\end{eqnarray} 
and the volume averaged velocity
$\mathbf{v} = \varphi \mathbf{v_s} + (1-\varphi) \mathbf{v_f}$ thus
follows the condition $\nabla. \mathbf{v} = 0$. The flux of colloids classically writes:
\begin{eqnarray}
\varphi \mathbf{v_s} = \varphi \mathbf{v} - D(\varphi) \nabla \varphi\,. \label{eq:colltrans}
\end{eqnarray}

In the confined geometry depicted in Fig.~\ref{fig:setup}, we assume that the meniscus can recede freely, and that colloids do not adsorb on the cell surfaces or migrate at the air/dispersion interface. 
The evaporative flux at the drying interface is not affected by the colloids as the  chemical activity of water remains always close to unity, even for strongly interacting small colloids (see e.g. Ref.~\cite{Daubersies:11}). Equation~(\ref{eq:sechage}) therefore still describes the drying kinetics of the drop. 

Axis-symmetry and the incompressibility condition $\nabla. \mathbf{v} = 0$ impose that 
the conservation of colloids follows:
\begin{eqnarray}
&&\frac{\partial \varphi}{\partial t }+ \mathbf{v}.\nabla \varphi = \frac{1}{r}\frac{\partial}{\partial r} \left( r D(\varphi) \frac{\partial \varphi}{\partial r}\right)+\frac{\partial}{\partial z} \left(D(\varphi) \frac{\partial \varphi}{\partial z}\right)\,. \label{eq:TransportConv}
\end{eqnarray}
For negligible convection within the drop, solvent evaporation from the drying interface is expected to induce  the formation of concentration gradients on a scale $\xi \sim D(\varphi)/\dot{R}$ due to the competition between collective diffusion and the receding of the meniscus~\cite{Daubersies:11}. However, these radial concentration gradients $\partial_r \varphi$ are associated to density gradients orthogonal to the gravity $\partial_r \rho(\varphi)$, which may in turn induce buoyancy-driven flows along $r$ (at least for a liquid dispersion). Such evaporation-induced natural convection has been 
reported in similar confined geometries and for various molecular mixtures~\cite{Selva:12,Daubersies:12,Lee:14}. 

We show recently in the case of weak concentration gradients along the drop (large $D(\varphi)$ and/or low $\dot{R}$), that these buoyancy-driven flows have no influence on the concentration gradients that generate them in a confined geometry~\cite{Selva:12}. This striking regime occurs for moderate Rayleigh numbers $\text{Ra} = v_m h/D \sim \mathcal{O}(1)$, where 
$v_m$ is the scale of the buoyancy-driven flows. Indeed, solutes (nanoparticles in our case) are convected by natural convection only on a scale $h \ll R$ during $\tau_d \sim h^2/D$ for $\text{Ra} \sim \mathcal{O}(1)$. As diffusion homogenizes concentrations over $h$ for $t\geq \tau_d$, and since the height-averaged radial component of the velocity is strictly~0 (mass conservation $\nabla. \mathbf{v} = 0$), the solute transport is dominated by diffusion. Averaging Eq.~(\ref{eq:TransportConv}) over the height $h$  for $\text{Ra} \sim \mathcal{O}(1)$ indeed leads to
\begin{eqnarray}
&&\frac{\partial \varphi}{\partial t } = \frac{1}{r}\frac{\partial}{\partial r} \left( r D(\varphi) \frac{\partial \varphi}{\partial r}\right)\,, \label{eq:Transport}
\end{eqnarray}
with the following boundary condition for the colloid flux at the receding meniscus: 
\begin{eqnarray}
- \left(D(\varphi) \nabla\varphi\right)_{r=R} = \varphi(r=R)\dot{R}\,. \label{BC1} 
\end{eqnarray}
In this regime of moderate Ra, concentrations are homogeneous over the cell height $h$ and mass transport is dominated by diffusion despite 
the presence of buoyancy-driven flows, see Ref.~\cite{Selva:12} for further details. This regime has been reported for several binary mixtures, including molecular solutions~\cite{Selva:12} and polymer solutions~\cite{Daubersies:12}. We will demonstrate later that it also occurs for the charged colloidal dispersion investigated in the present work.
Finally, the average concentration within the drop $<\varphi>$ is simply related to the normalized area $\alpha$ through:
\begin{eqnarray}
<\varphi> = \frac{1}{\pi R^2}\int_0^R 2\pi r \varphi(r,t) \text{d}r = \frac{\varphi_0}{\alpha} \label{eq:phimoy}
\end{eqnarray}

In the confined drop, solvent evaporation induces the formation of concentration gradients at the drying interface on a scale 
$\xi \sim D(\varphi)/\dot{R}$. For radii $a = 10~$nm, Stokes-Einstein estimate $D_0 \approx 2 \times 10^{-11}$~m$^2$/s, predicts the formation of a thin crust during the drying kinetics: $\xi \simeq 200~\mu$m for $\dot{R} \simeq 100$~nm/s. 
As shown later, our experimental results for charged dispersions of small colloids show that
concentration gradients over the drop are much more weaker, and associated to collective diffusion coefficients
$D(\varphi) \approx 10$--$30 D_0$ for the whole range of concentration. Such high values prevent from the formation of a thin crust, and lead only to slight concentration gradients.    
Moreover, our experiments reveal a transition from a liquid dispersion (sol) to a poro-elastic solid  (gel) at $\varphi \approx 0.3$.
Below  $\varphi \approx 0.3$, the weak concentration gradients induce natural convection as explained above, but the 
transport remains dominated by diffusion as  $\text{Ra} \sim \mathcal{O}(1)$, see Sec.~\ref{sec:gel}.
We will also discuss in Sec.~\ref{sec:discussion} how the description of the collective diffusion coefficient related to the osmotic compressibility of the dispersion and the permeability to the relative flow solvent/colloids~\cite{Russel,Nagele} is affected by such a liquid$\to$solid transition~\cite{Style:11}.

\section{Materials and Methods}
\subsection{Confined drying experiments}
We use glass wafers (3'' diameter, 1~mm thickness) coated by a thin layer of cross-linked poly(dimethylsiloxane) PDMS 
(thickness $\simeq 30~\mu$m, Sylgard 184). The thin PDMS layer prevents from the pinning of the receding meniscus, at least in the first stage of the drying process, see Sec.~\ref{sec:Results}.
We pay a particular attention to the neatness of the PDMS layer, as any adsorbed dust may pin the receding meniscus and lead to mechanical instabilities (see Sec.~\ref{sec:Results}).
The spacers consist of small pieces ($\simeq 6$~mm$^2$) made of glass or PDMS films, with thicknesses $h$ ranging from 80 to 250~$\mu$m. Room humidity ranges typically from $a_e=0.4$ to 0.65,
 with variations below 0.05 during a given experiment.

In a classical experiment, we place a drop with a controlled volume of $\simeq 1~\mu$L at the center of one of the glass wafers. We then carefully close the cell
using the second wafer and start immediately a video acquisition using a stereomicroscope and a CCD camera. Classical resolution is 5~$\mu$m/pixel and 
frame rate is 6~images/min. 

To visualize flow patterns within the drop, we seed the colloidal dispersion with a small amount of fluorescent particles (Fluorospheres, diameter 500~nm, carboxylate-stabilized, volume fraction $<0.01\%$). Trajectories are recorded using fluorescence imaging using the same stereomicroscope as above. To measure precisely flow profiles, see Fig.~\ref{fig:Tracking}, we use an inverted microscope (IX71, Olympus) and a high NA objective (60X, NA$=1.2$)
to monitor the trajectories of the tracers  at different focal planes with steps of 10~$\mu$m or 20~$\mu$m   
(the typical field of view of the camera is $\ll R$), see Fig.~\ref{fig:setup}(b). Standard tracking algorithms are used to compute the velocities of the tracers as a function of the height $z$  within the cell.

\subsection{Dispersions \label{Sec:dispersion}}  
We investigate Ludox dispersions of silica nanoparticles, and more precisely the monodisperse anionic grade AS40  
(Sigma-Aldrich).
The mean colloid radius is $a = 11~$nm, the mass fraction of the stock dispersions is $\simeq 40\%$ and these colloids are stabilized by negative silanol groups at their interface. The surface charge density depends on the pH of the dispersion (pH$=9.2$) and on  the counterions concentration, and is typically of 0.5~e/nm$^2$~\cite{Jonsson:11,Persello}. 
To estimate the volume fraction $\varphi_0$ of the dispersion, we measure its density and its dry extract 
(at 150$^\circ$C for $\simeq 30$~min), and we assume the additivity of the volumes.  These measurements lead to 
the density of the silica particles $\rho_s = 2.216$~g/mL and $\varphi_0 \simeq 0.24$. 

The aim of the present work is  to demonstrate that electrostatic interactions play a large role for the drying process, 
even for  stock dispersions which may thus contain ionic species. 
In our experiments, we mainly use AS40 dispersions as received and we thus do not control their exact ionic content by using osmotic equilibrium against known solutions.
The counterions of  AS40 dispersions are ammonium hydroxide ions, as compared to the other  monodisperse anionic grade TM40 for which counterions are sodium ions~\cite{Grace}. AS40 dispersions are often used in applications for which the presence of
sodium is detrimental. Ammonium hydroxide ions are indeed in equilibrium with ammonia gas dissolved in the dispersion, and the later may eventually evaporate~\cite{Grace}.  To check whether our results depend on the possible counterions volatility or not, we also investigated TM40, and we observe similar behaviors but with 
slightly shifted concentration values (data not shown).

\subsection{Raman micro-spectroscopy}
We use a custom-made Raman micro-spectrometer setup coupled to an inverted microscope (Olympus IX71 and Andor Shamrock).
A laser beam ($\lambda=532~$nm, Coherent Sapphire SF) is focused with a 20X objective (NA$=0.45$) at approximatively $z \approx h/2$ within the cell. Scattered light is collected by the same objective, filtered to get the Raman contribution, 
and directed to the spectrometer, see Fig.~\ref{fig:setup}(b). 
A confocal pinhole ($100~\mu$m) conjugated with the focal plane prevents from collecting excessive out-of-focus contributions. Typical experimental parameters are: acquisition time 2~s, slit~$200~\mu$m, laser power at the focal plane $\simeq 25$~mW, and grating 600~lines/mm.

To monitor the concentration process in space and time during drying, we scan the drop along its diameter ($2R_0 \simeq 2$~mm)
using a motorized stage (M\"arzh\"auser) synchronized with the Raman measurements. 
A typical spatial scan lasts 2~min with a spatial step of 30~$\mu$m.
This process is repeated up to the complete consolidation in order to get space-time Raman measurements of the drying dynamics.

\begin{figure}[htbp]
\begin{center}
\includegraphics{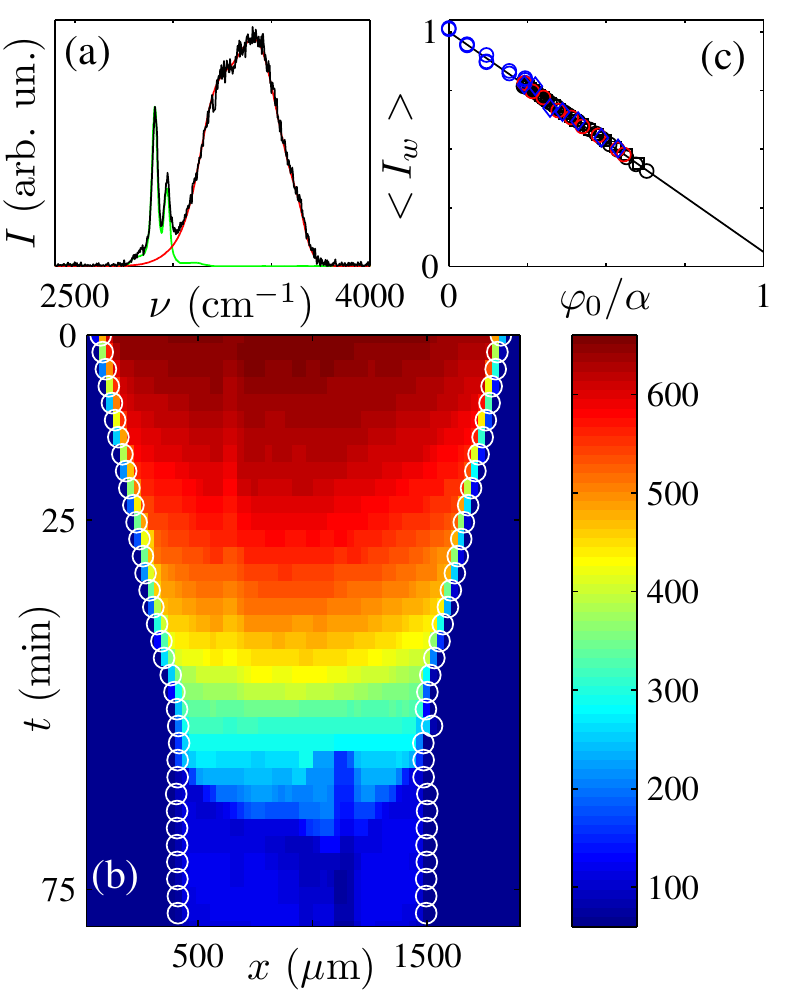}
\caption{(Color online) (a) Typical Raman spectrum acquired within the drop (black). The green and red spectra correspond to the Raman contribution of PDMS and water respectively. (b) Typical space-time plot of the water contribution during the drying of a drop ($h=250~\mu$m, drop volume 1~$\mu$L). The white circles are  the positions of the drop meniscus.
(c) Different calibration curves showing the drop-averaged water contribution $<I_w>$ vs. the averaged colloid concentration within the drop
$\varphi_0/\alpha$ (black symbols). The colored symbols correspond to other measurements obtained with 60X (blue) and 20X (red) objectives  at steady state or by simple dilution (see text). The continuous line is the linear fit of the data used to convert the Raman signal into a colloid concentration $\varphi$.}  
\label{fig:Calib}
\end{center}
\end{figure}

\subsection{Raman calibration and screening of the drying kinetics \label{sec:Ramantech}}

Figure~\ref{fig:Calib}(a) shows a typical Raman spectrum acquired within the drop.
This spectrum is corrected for the baseline measured in the range 2400--2700 and 3850-4000~cm$^{-1}$.
In the investigated spectral range, the data evidence a  contribution  of the PDMS layers (2800--3000~cm$^{-1}$) despite the confocal pinhole, and  a wide contribution in the region 3500~cm$^{-1}$ due to the OH stretching and bending
modes of the water molecules.  The measurements of the silica concentration from such data are not straightforward as the Raman spectra do not display any signature of the silica particles in the spectral range investigated~\cite{ALESSI2013}.
However, we show below using a careful calibration that quantitative measurements of the colloid volume  fraction $\varphi$ are still possible from the measurements of the Raman water contribution only.

Figure~\ref{fig:Calib}(b) shows a typical space-time plot of the Raman water contribution (estimated from the spectra at 3400~cm$^{-1}$) during the drying of drop. As  kinetics are rather long 
($\simeq 1$~h), we ignore the small acquisition time of a spatial scan to build this space-time plot.
The meniscus positions are determined with a subpixel resolution from such data, and the dynamics $R$ vs. $t$ is thus known.  
We only consider the first 55~min of the data (before the final consolidation with a steady radius $R$), and we do not 
considered the Raman data acquired at longer time scales as the drop delaminates from the wafers and cracks appear, see Sec.~\ref{sec:Results}.

The mean water contribution within the drop decreases from $\simeq 650$ to $\simeq 260$ (arbitrary units) at $t\simeq 55$~min for this data set with slight spatial gradients  over the drop  ($<6~\%$), all along the  drying process.
These measurements provide a self-calibration as the  drying kinetics $R$ vs. $t$ leads to the temporal evolution of the average concentration $<\varphi> = \varphi_0 / \alpha$, see Eq.~(\ref{eq:phimoy}). Figure~\ref{fig:Calib}c shows the average water contribution 
$<I_w> = (2/R^2)\int_0^R \text{dr} \, r I_w(r,t)$ as a function of $\varphi_0 / \alpha$. These measurements, normalized to get $<I_w>\to 1$ when $<\varphi> \to 0$, clearly evidence the affine behavior $<I_w> = 1-0.94<\varphi>$. 
Figure~\ref{fig:Calib}(c) displays different superimposed data sets evidencing the reproducibility of this relation. 

To check the robustness of this calibration, we also performed independent measurements using both the 20X objective, and an oil-immersion objective (60X, NA$=1.4$) at a focal plane located $\simeq 50~\mu$m from the bottom PDMS layer. To control the concentration within the drop, we proceeded as follows. We first put diluted drops of known concentrations and we acquire Raman spectra before significant drying.
The corresponding data follow again the affine behavior (colored symbols, Fig.~\ref{fig:Calib}(c)). Then, we put a drop of the AS40 dispersion 
($\varphi_0 \simeq 0.24$), we measure its initial area, and we let the solvent evaporates for $\simeq 20$~min. We then stop evaporation by covering the cell by a closed box containing  suspended water drops to saturate the cell with a high humidity, and hence stopping the drying kinetics. To make sure concentration gradients along the drop relax through collective diffusion,
Raman spectra are acquired $\simeq 10$~min after evaporation stops. The area of the drop is again measured to estimate the concentration $\varphi = \varphi_0/\alpha$
within the drop. This process is repeated until the drop finally consolidates. 
These measurements (colored symbols,  Fig.~\ref{fig:Calib}(c))  again confirmed that the (normalized) Raman water contribution follows the same affine law. 

We expected the linear behavior $<I_w> = 1-<\varphi>$ as (i) Rayleigh scattering does not contribute to the measured signal for such small particles (silica refractive index $\simeq 1.46$, $a\simeq 11$~nm), and (ii) as the Raman signal is a priori proportional to the number of water molecules contained within the sample volume. 
The observed affine relation $<I_w> = 1-0.94<\varphi>$ suggests that a water contribution of $\simeq 0.06$ remains even when
$\varphi \to 1$. This small contribution may be due to an overestimation of the volume fraction  $\varphi_0$
using our dry extract measurements (see Section~\ref{Sec:dispersion}),
as water molecules may remain attached to the silica surface despite this thermal treatment 
(as recently discussed  by Piroird {\it et al.} \cite{Piroird:16}). We do not investigate this issue in  more details, and we use the calibration  displayed in Fig.~\ref{fig:Calib} to extract the volume fraction $\varphi$ from the measured Raman water contribution.

\section{Results \label{sec:Results}}

\subsection{Drying kinetics and observed phenomenology}

\begin{figure}[htbp]
\begin{center}
\includegraphics{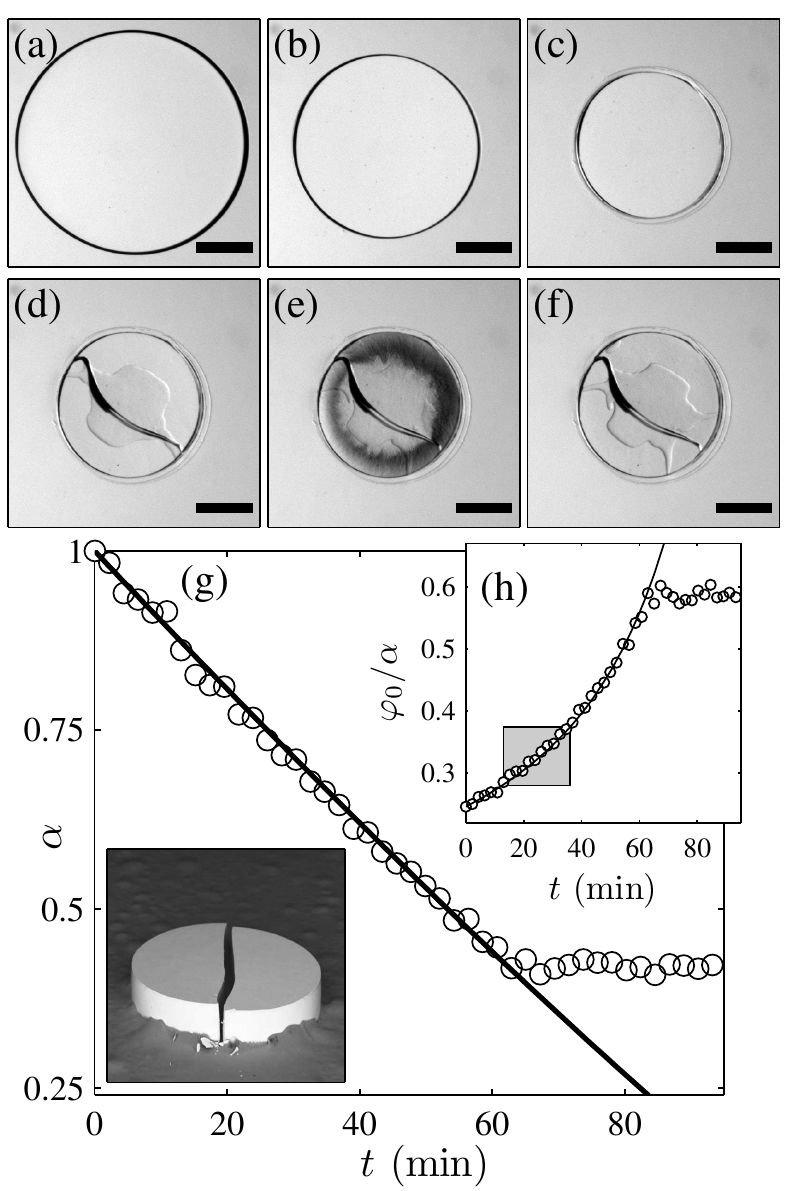}
\caption{(a)-(f) Typical bright field observation of the drying dynamics for $t=0$, $34$, $59$, $62$, $67$ and $92$~min 
(scalebar $500~\mu$m, $h=250~\mu$m, see also the corresponding movie M1.avi~\cite{SM2016}).
On snapshot (c), the receding meniscus leads to a deposit, and the material further delaminates and fractures (d). Air finally 
invades the pores of the solid network (e), up to the final material (f) (see also the inset of (g) for a SEM view).
(g) Normalized area $\alpha$ vs. time estimated from the analysis of the drying dynamics (see text). (h) Corresponding average concentration $\varphi_0/\alpha$ vs. time $t$. The gray zone corresponds to the time interval at which the {\it gelation front} crosses the drop (see Fig.~\ref{fig:Lines} and Section~\ref{sec:gel}).
\label{fig:bino}}
\end{center}
\end{figure}

Figure~\ref{fig:bino} shows typical bright field observations of the drying of a AS40 dispersion drop at $\varphi_0 \simeq 0.24$.
At early time scales ($t<60$~min), the drop remains circular and its radius decreases smoothly. The meniscus recedes freely on the PDMS surface and no deposit is formed. When the PDMS layers are not carefully cleaned, the observed scenario is quite different (see the movie M5.avi in Supplemental Material~\cite{SM2016}). The  receding contact line may indeed pin at a given location, leading to the further buckling of the drop (often 
referred to as the {\it invagination} instability). Such behaviors were  reported  earlier by Pauchard {\it et al.} in
similar experiments and such instabilities were hindered by using glass wafers coated 
by liquid lubricating layers~\cite{Boulogne:13}. In the following, we only focus on experiments with clean wafers and the drops remain circular during drying.

From image analysis, we extract the normalized drop area $\alpha$ vs.  $t$ and thus the mean concentration $\varphi_0/\alpha$ within the drop. 
The temporal evolutions $\alpha(t)$ are nicely fitted by Eq.~(\ref{eq:tauf}) with values for $\tilde{D}$ ranging from 2.3 to $3 \times 10^{-10}$~m$^2$/s for several experiments. 
At $t\approx 60~$min, and thus $\varphi \approx 0.6$, a deposit forms on the PDMS layers and a sequence of mechanical instabilities rapidly occurs, see Fig.~\ref{fig:bino}. The drop first delaminates from the PDMS layers, and a crack often appears across the drop. 
At longer time scales, a sharp change of refractive index suggests that air invades the drop from its outer boundary (e). Ultimately, the final material does not evolve anymore (f), and the corresponding SEM images show materials with sharp interfaces
(see the inset of Fig.~\ref{fig:bino}(g)). We do not focus in the present work on such mechanical instabilities, and we only investigate the early stage of drying  corresponding to the concentration process up to the final consolidation ($t\leq 60~$min).

\subsection{Concentration profiles and collective diffusion coefficient} 

\begin{figure}[htbp]
\begin{center}
\includegraphics{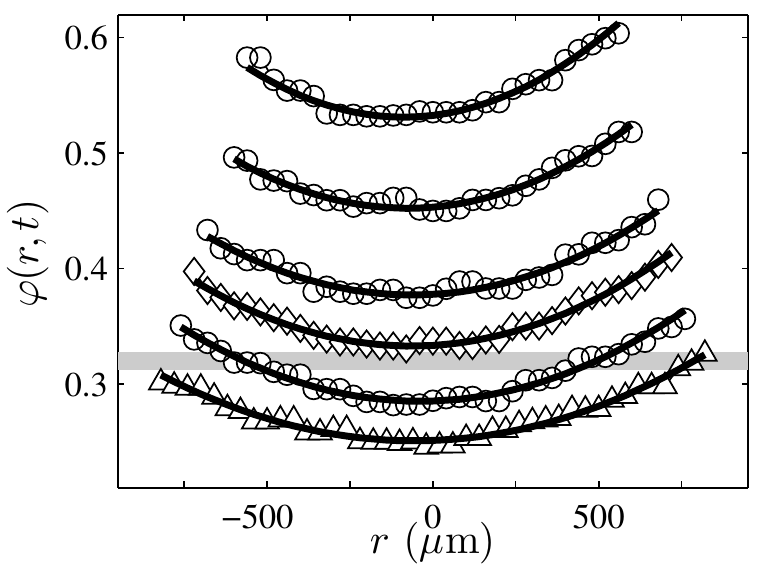}
\caption{Typical concentration profiles $\varphi(r,t)$ measured within the drop using Raman spectroscopy ($h=170~\mu$m, drop volume 0.75~$\mu$L). The continuous lines
are parabolic fits, see Eq.~(\ref{eq:parabole}). $\triangle$ ($\diamond$ resp.): concentration profile  corresponding to the 
start (the end resp.) of the drop gelation (see Section~\ref{sec:gel}). The gray line is $\varphi_c \simeq 0.32$.}  
\label{fig:Profil}
\end{center}
\end{figure}

Figure~\ref{fig:Profil} shows a typical temporal evolution of the colloid concentration field measured using our Raman technique, see Section~\ref{sec:Ramantech}. 
We plot in Fig.~\ref{fig:Profil} only a few curves along the drying process for the sake of clarity (the temporal resolution of the full data is $\simeq 2~$min, see Fig.~\ref{fig:Calib}(b)). 
These concentration profiles clearly reveal that evaporation does not lead to the formation of a thin crust of 
concentrated colloids at the receding boundary, but 
only to slight gradients along the drop. 
These profiles are well-fitted by 
\begin{eqnarray}
\varphi(r,t) = \varphi_c(1+\epsilon r^2/R^2)\,, \label{eq:parabole}
\end{eqnarray} 
where $\varphi_c$ is the concentration at $r=0$ and $\epsilon \ll 1$, see the continuous lines in Fig.~\ref{fig:Profil}.

Assuming that the description of the transport process within the dispersion is correct, see Sec~\ref{sec:confineddrying} and namely Eqs.~(\ref{eq:Transport}-\ref{BC1}), we can now estimate the
collective diffusion coefficient $D$ vs. $\varphi$ from such measurements. Indeed, both concentration gradients $\partial_r \varphi$ and 
drying kinetics $\dot{R}$ are measured, and  the boundary condition~Eq.~(\ref{BC1}) possibly gives  an estimate of $D(\varphi)$.  
To minimize the dispersion of $D(\varphi)$ induced by the estimation of two numerical derivatives ($\partial_r \varphi$ and $\dot{R}$), $R$ vs. $t$ is fitted using Eq.~(\ref{eq:tauf}) to estimate precisely $\dot{R}$ (see Fig.~\ref{fig:bino}(g)), and the fits of the profiles by Eq.~(\ref{eq:parabole}) yield estimates of the gradients at $r=R$ (see Fig.~\ref{fig:Profil}).

\begin{figure}[htbp]
\begin{center}
\includegraphics{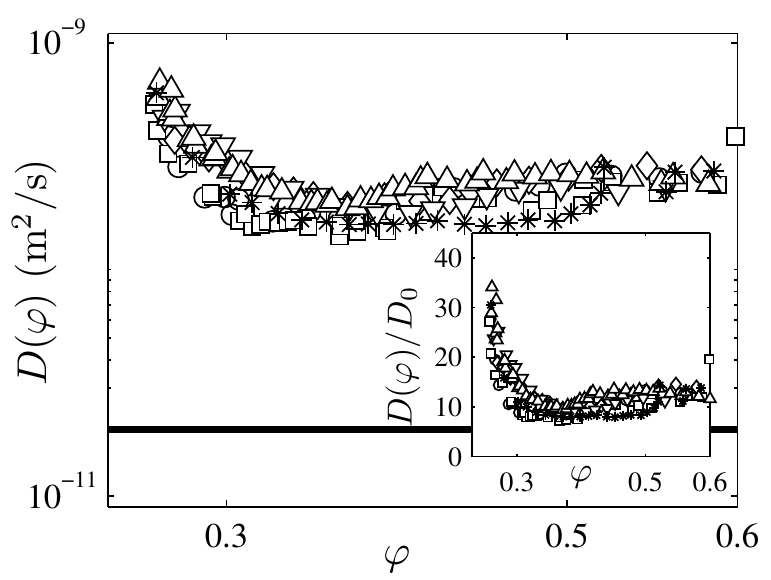}
\caption{Collective diffusion vs. concentration estimated using the combined measurements of the drying dynamics $R(t)$ and the concentration profiles $\varphi(r,t)$, see Eq.~(\ref{BC1}). The different symbols correspond to experiments performed with cells of different heights $h=80$, 170 and 250~$\mu$m.   The continuous line is the Stokes-Einstein estimate $D_0 \simeq 2 \times 10^{-11}$~m$^2$/s. Inset: same data normalized by $D_0$ 
in a linear-linear plot. }
\label{fig:CollDiffusion}
\end{center}
\end{figure}

Figure~\ref{fig:CollDiffusion} shows the output of such measurements for various cell heights ranging from 80 to 250~$\mu$m. 
For the whole concentration range, $D(\varphi)$ deviates significantly from the Stokes-Einstein estimate $D_0 =2\times 10^{-11}$~m$^2$/s as $D(\varphi) \simeq 10$--$30 D_0$, evidencing the crucial role played 
by the electrostatic interactions on the relaxation of concentration gradients, even for such a stock dispersion
(i.e. without removing ionic species). 
Despite the spread of the data related to the estimate of two numerical derivatives, our measurements outline the following behavior:  $D(\varphi)$ first decreases from $\simeq 6$ 
to $\simeq 2 \times 10^{-10}$~m$^2$/s for $\varphi$ increasing from $\varphi_0$  to $\simeq 0.35$, 
and $D(\varphi)$ remains nearly constant $\simeq 2 \pm 0.5 \times 10^{-10}$~m$^2$/s for larger $\varphi$.

Boulogne {\it et al.} performed similar experiments as those reported in Fig.~\ref{fig:bino} using different charged dispersions, including  
Ludox HS40 and TM50  with volume fractions $\varphi_0$ ranging from 0.1 to 0.22~\cite{Boulogne:13}. 
Their observations are quite different, as they always reported invagination of the drop, suggesting the formation of an elastic crust at the outer of the drop and its buckling. These observations were not linked to the pinning of the receding meniscus, as the contact line receded freely on thin viscous lubricating layers on the wafers. However, their  different geometry (larger drops $R_0\approx 2.9$~mm, and smaller wafers $R_w=8~$mm), leads to significantly larger velocities  $\dot{R}\simeq 400~$nm/s (see Fig.~2 in Ref.~\cite{Boulogne:13}). If we assume that our measurements of $D(\varphi)$ apply to their investigated dispersions, the expected scale of the concentration gradient is $\xi = D(\varphi)/\dot{R} \simeq 250~\mu$m $ \ll R_0$, whereas $\xi \sim R_0 \simeq 1~$mm in our case. This may  confirm the formation of a crust of concentrated colloids at the receding meniscus in their experimental configuration.

\subsection{Flow patterns: from drying-induced buoyancy flows to homogeneous drying \label{sec:gel}}

To validate the use of Eq.~(\ref{BC1}) to estimate $D(\varphi)$, we should also confirm that the colloid transport is dominated by collective diffusion even if natural convection occurs within the drop, see Eq.~(\ref{eq:Transport}). 
We thus seed the dispersion using fluorescent tracers and monitor their trajectories during drying, see Fig.~\ref{fig:Lines}(a-b). 
These experiments reveal two distinct flow patterns, see the movie M2.avi in Supplemental Material~\cite{SM2016} and the space time plot in Fig.~\ref{fig:Lines}(c) along a given drop radius. 
At early stages (typically $t \leq 15~$min), we observe axis-symmetrical recirculating flows  which convect the tracers, from the meniscus to the drop center for $z<h/2$ and from 
the drop center to the meniscus for $z>h/2$. The maximal radial velocity of such flows is of the order of 1~$\mu$m/s for $h=250~\mu$m. These flows then stop in the range $t = 15~$--30~min, and the tracers follow further radial trajectories 
towards the center of the drop, uniformly over the cell height. In this last regime, tracers concentrate homogeneously, leading us to refer to this regime as the {\it homogeneous drying} regime.

As discussed earlier, the axis-symmetrical flows observed at early time scales correspond to buoyancy-driven flows, as also 
recently demonstrated using similar experiments but with molecular mixtures~\cite{Selva:12,Daubersies:12,Lee:14}. The  {\it homogeneous drying} regime suggests the transition to a solid in which tracers are trapped. Despite the obvious 
displacements of fluorescent tracers during the drop drying, we also show below that our measurements of collective 
diffusion coefficients remain valid, as the mass transport  is still described by Eq.~(\ref{eq:Transport}).

\begin{figure}[htbp]
\begin{center}
\includegraphics{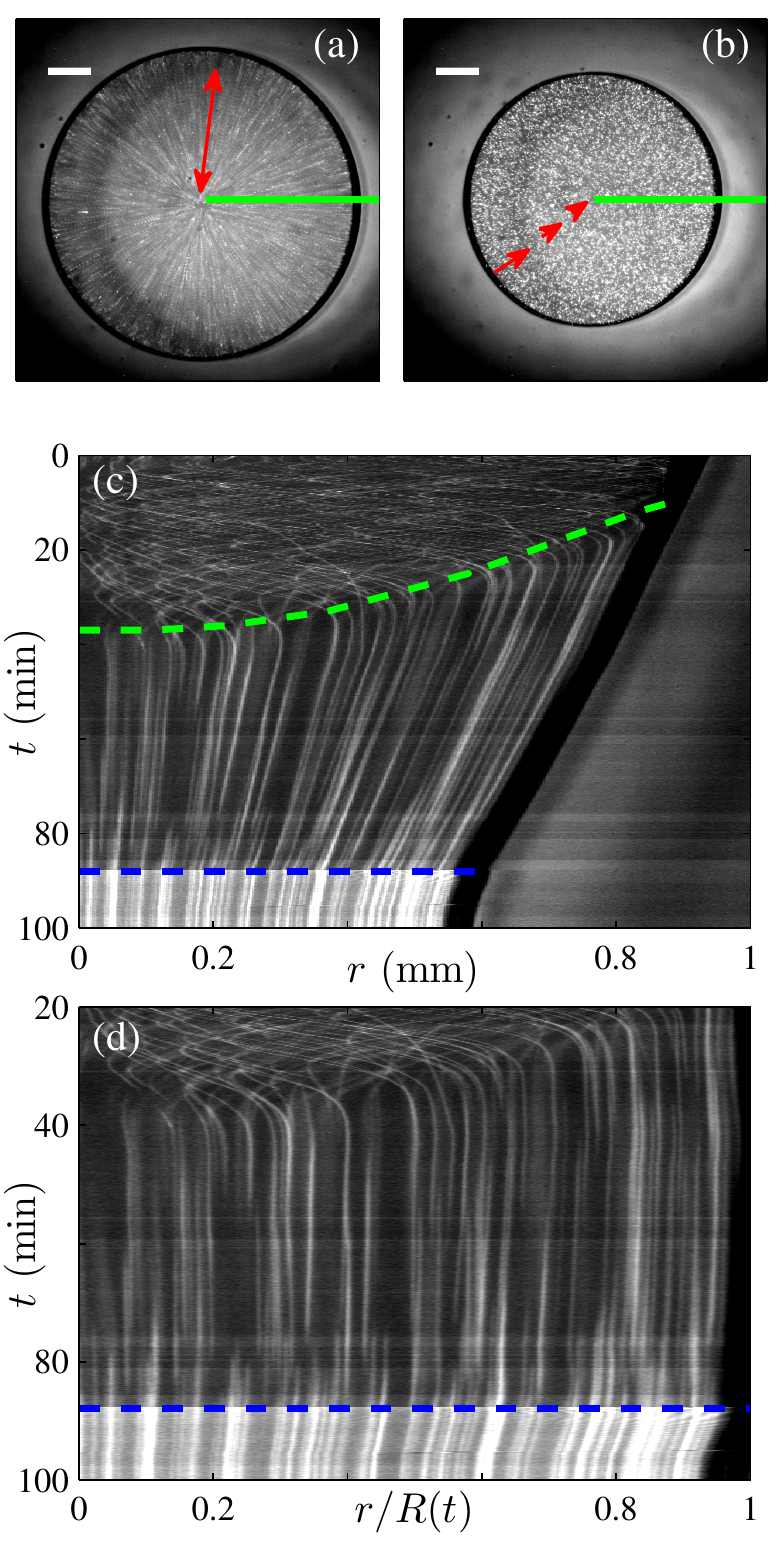}
\caption{(color online) (a) and (b) Fluorescence images obtained by superimposing 10 successive frames at time $t=4$~min (a) and $t=50$~min (b) corresponding to the two regimes discussed in the text 
(scalebar $250~\mu$m, $h=250~\mu$m).
The  streaks  help to highlight the flow within the drop, see also the movie M2.avi~\cite{SM2016}. The red arrows show the two different flow patterns.
(c) Space-time plot of the fluorescence intensity along the green lines shown in (a) and (b). (d) The same space-time plot, but re-scaled  with $r/R(t)$  (see also the movie M3.avi~\cite{SM2016}). 
\label{fig:Lines}}
\end{center}
\end{figure}

\subsubsection{Buoyancy-driven flows}
The measurements of the concentration profiles shown in Fig.~\ref{fig:Profil} are associated to density 
gradients along the drop. For a liquid dispersion, these weak concentration gradients induce natural convection which can be  calculated using 
the lubrication approximation. Assuming that the variation of the viscosity $\eta$ of the dispersion  along the 
drop is negligible, the radial velocity component is given by~\cite{Selva:12}:
\begin{eqnarray}
&&v_r(z) = \frac{(\rho_s-\rho_e) g h^3 \partial_r\varphi}{12 \eta}\,\tilde{z}(1-\tilde{z})(2 \tilde{z}-1) \label{eq:buoyancy}\,, 
\end{eqnarray} 
with $\tilde z=z/h$, and $\rho_e$ the water density.
The maximal flow velocity (for $\tilde z \simeq 0.2$ and $\tilde z \simeq 0.8$) is roughly given by $v_m \simeq 0.008 \delta\rho g h^3/(\eta R)$, where $\delta\rho$ is the density difference across the drop.

\begin{figure}[htbp]
\begin{center}
\includegraphics{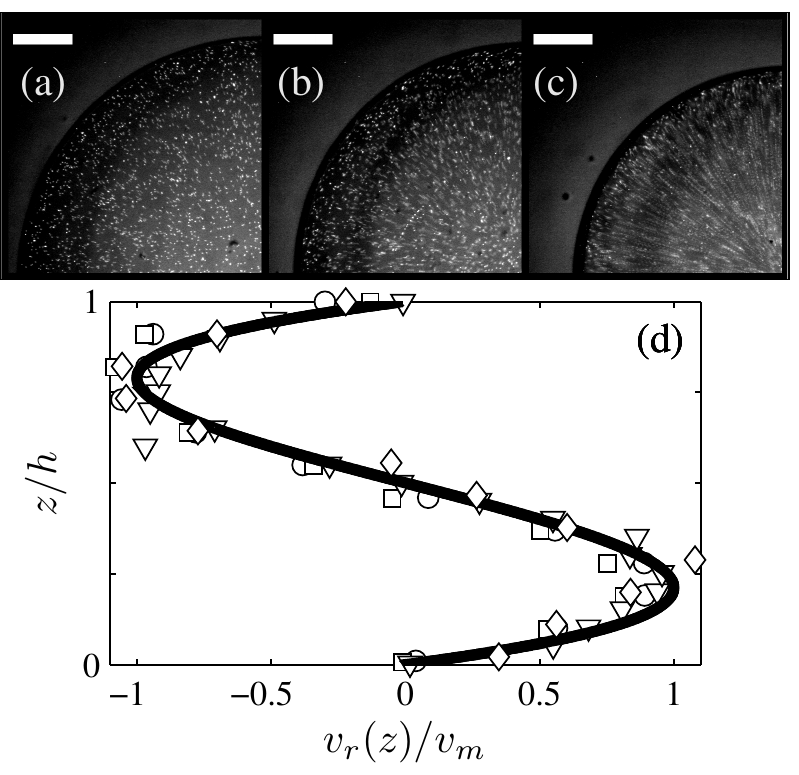}
\caption{Top: these pictures show the averaging of 4 successive fluorescence images to evidence flows within the drop, for different cell heights (a) $h=80$, (b) 170 and (c) 250~$\mu$m.
Only a drop quarter is shown for clarity, scale bars 250~$\mu$m, see also the movie M4.avi in Supplemental Material~\cite{SM2016}. 
(d) Velocity profiles measured using particle tracking in the drying-induced buoyancy regime for $h=170$ and $250~\mu$m at $r=R(t)/2$, and normalized by the maximal velocity $v_m$. The continuous line is the theoretical prediction given by
Eq.~(\ref{eq:buoyancy}). The different symbols correspond to different measurements.
}  \label{fig:Tracking}
\end{center}
\end{figure}
Figure~\ref{fig:Tracking}(d) shows velocity profiles measured using particle tracking for different drops ($h=170$ and 250~$\mu$m) and at $r=R(t)/2$ along the cell height $h$. These profiles are well fitted by Eq.~(\ref{eq:buoyancy}) validating the above description. More importantly, Figs.~\ref{fig:Tracking}(a--c) report streak-like velocimetry
to evidence the strong dependence of these  buoyancy-driven flows with the cell height $h$ 
($v_m \sim h^3$, see also the corresponding movie in SI). For $h=250~\mu$m,
velocities $v_m$ are of the order of 1~$\mu$m/s at the very beginning of the drying process, 
and smaller than  0.4~$\mu$m/s for $h=170~\mu$m, and even smaller than 100~nm/s for $h=80~\mu$m. These 
values are in rough agreement with the theoretical estimates given by Eq.~(\ref{eq:buoyancy}) and the viscosity of the stock dispersion measured using classical rheometry ($\eta \simeq 15~$mPa.s, cone-plane geometry, Kinexus Malvern rheometer) leading to $v_m \simeq 0.6~\mu$m/s for $h=250~\mu$m and a gradient $\partial_r \varphi \simeq 0.06/R$ at $r=R/2$ (see the first concentration profiles shown in Fig.~\ref{fig:Profil}). Our estimations should be taken with care considering that the 
viscosity of the dispersion evolves with the volume fraction both in space and time. Moreover, Eq.~(\ref{eq:buoyancy}) is strictly valid within the lubrication approximation which does not strictly apply to our investigated geometries ($R_0~\simeq 1$~mm, $h$ up to 250~$\mu$m).

As discussed earlier in this work, these buoyancy-driven flows may have no influence on the concentration gradients that generate them in a confined geometry. This regime occurs when the Rayleigh number $\text{Ra} = v_m h/D$, follows  $\text{Ra} \sim \mathcal{O}(1)$, see Ref.~\cite{Selva:12} for more details. 
In our experiments, $D(\varphi)$ ranges  from $6 \times 10^{-10}$~m$^2$/s to $2\times 10^{-10}$~m$^2$/s for $\varphi = 0.24$--0.35,
see Fig.~\ref{fig:CollDiffusion}. $\text{Ra}$ thus ranges from 0.8 to 2.5 for $h=250~\mu$m and $v_m \simeq 1~\mu$m/s (i.e. $\text{Ra} \sim \mathcal{O}(1)$), demonstrating that the drying-induced concentration gradients in silica nanoparticles is not affected by the buoyancy-driven flows. These flows, however, can convect larger colloidal species such as the fluorescent tracers dispersed in the drop.
 
This result is even confirmed directly
by the measurements reported in Fig.~\ref{fig:CollDiffusion}. The values of $D(\varphi)$ indeed do not depend on $h$, whereas 
buoyancy-driven flows strongly depend on $h$ as $v_m \sim h^3$, see also Figs.~\ref{fig:Tracking}(a--c) and the corresponding movie in SI.

\subsubsection{Gelation}
At a given time, these buoyancy-driven flows suddenly vanish close to the receding meniscus $r \simeq R$, and this 
phenomenon progresses within the drop up to its center. At longer time scales, the fluorescent tracers seem to be trapped within the dispersion, and they follow radial trajectories towards the center of the drop.

These observations suggest the formation of a gel which traps the fluorescent tracers, that further dries up to the
drop consolidation without sticking on the PDMS layers. The {\it gelation front} which crosses the drop from $r=R$ up to $r=0$ (see the green dotted line in Fig.~\ref{fig:Lines}(c)), suggests that this transition occurs at a critical concentration $\varphi_c$.

Indeed, our combined measurements of fluorescence imaging and concentration profiles make it possible to confirm this picture and to roughly estimate  $\varphi_c$. We first extract the normalized drop area $\alpha_1$ at which we observe the 
beginning of the gelation at $r=R$, and the normalized area  $\alpha_2$   at which the gel fully invades the drop (see 
the corresponding range highlighted by the green dotted line in ~Fig.~\ref{fig:Lines}(c)).
We then report in Fig.~\ref{fig:Profil}, the concentration profiles corresponding to these two critical $\alpha$. For  $\alpha \simeq \alpha_1$ (symbols $\triangle$), concentration reaches $\varphi_c \simeq 0.32$ at $r=R$, while for $\alpha \simeq \alpha_2$ (symbols $\diamond$), one observes $\varphi(r,t) \simeq \varphi_c$ at $r = 0$ (see the gray line in  Fig.~\ref{fig:Profil}). The crossing of the gelation front within the drop arises from the existence of slight concentration gradients which intersect the concentration $\varphi_c$ at different positions $r$, see e.g. the intermediate profile 
displayed Fig.~\ref{fig:Profil}.

Our measurements give a rough estimate $\varphi_c \simeq 0.32 \pm 0.02$ for this liquid$\to$solid transition which 
should be considered cautiously. Indeed, we estimate $\varphi_c$ from the vanishing of the buoyancy-driven flows, while the
latter may also vanish for a sudden increase of viscosity, see Eq.~(\ref{eq:buoyancy}). Nevertheless, 
this rough estimate is in good agreement  with other reported values in the literature for similar systems.  
Ludox dispersions, and more generally any dispersion of highly charged particles, are indeed well-known to form 
elastic gels at concentrations well-below close-packing. This feature has been 
reported many times in the context of drying, at volume fractions $\varphi$ ranging from 0.1 to 0.4 depending on the 
ionic content of the dispersion, and the surface charge density (in-situ SAXS , indentation, microscopy)~\cite{Boulogne:14,Ziane:15a,Bodiguel:10A}.
Direct rheological measurements were also reported on a similar Ludox dispersion (HS40) showing a sharp transition from a 
viscous dispersion (sol) to elastic gels at $\varphi \simeq 0.3$~\cite{DIGIUSEPPE2012}. 
We also confirmed our measurement $\varphi_c \simeq 0.32$ by performing slow drying experiments of AS40 dispersions in small vials (drying times $\simeq 30$--60~h at $T=40^\circ$C under constant stirring) as in Ref.~\cite{DIGIUSEPPE2012}. These experiments reveal the formation of elastic gels for concentrations roughly above $\varphi_c \simeq 0.3$ (measured using dry extract).  

To locate this liquid$\to$solid transition within the whole dynamics, we also report in Fig.~\ref{fig:bino}
the $\alpha$ range (and thus the temporal range) at which the gelation front crosses the drop. This plot evidences that
this transition occurs long before the final consolidation of the drop.

\subsubsection{Homogeneous drying}
We now investigate the trajectories of the tracers in this gel phase.
Figure~\ref{fig:Lines}(d) shows the same space-time plot as in (c), but against the rescaled variable $r/R(t)$. Trajectories 
in this space-time plot are vertical lines, suggesting that the radial velocities of the tracers are given by 
\begin{equation}
v_r \simeq \frac{r}{R} \dot{R}\,, \label{eq:traj}
\end{equation}
without any variation over the cell height.
This feature is better evidenced using the movie M3.avi in Supplemental Material~\cite{SM2016}, for which a zoom is continuously  adjusted to get a fixed drop radius on the screen. In this movie, tracers seem immobile showing  that they homogeneously concentrate within
the drop.

These results enable us to assess that the trajectories of the fluorescent tracers in this second regime do not follow
the volume-averaged velocity $v = \varphi v_s + (1-\varphi) v_f = 0$, see Sec.~\ref{sec:confineddrying}, but more likely the velocity of the silica nanoparticles $v_s$. Tracers are indeed trapped within the colloidal elastic network, and follow its deformation. 
Relation~(\ref{eq:traj}) simply shows that the drop dries {\it homogeneously}, i.e.
$\varphi(r,t) \approx \varphi_0 R_0^2/R^2$, as the concentration gradients are of the order of $\epsilon \ll 1$, see Eq.~(\ref{eq:parabole}) and Fig.~\ref{fig:Profil}. 

The same result can be demonstrated more rigorously using the framework of large deformation poro-elasticity~\cite{MacMinn2016}.  One can indeed relate the (Eulerian) radial displacement field $u_s$ of the solid network to the local concentration field $\varphi(r,t)$. This relation takes the following form in our confined cylindrical geometry:
\begin{eqnarray}
u_s = r -\left( r^2 + 2 \int_0^r (\frac{\varphi}{\varphi_0}-1)r \text{d}r \right)^{1/2}\,, \label{eq:phidonneu}
\end{eqnarray}
see for instance Eq.~(14) in Ref.~\cite{Bertrand2016} for a demonstration in the case of a spherical geometry.
Assuming in the above equation that $\varphi(r,t) \approx \varphi_0 R_0^2/R^2$ at first order, leads to:
\begin{eqnarray}
u_s \approx r - r\frac{R_0}{R}\,,
\end{eqnarray}
and finally to
\begin{eqnarray}
v_s = \frac{\partial_t u_s}{1-\partial_r u_s} \approx \frac{r\dot{R}}{R}\,.
\end{eqnarray}
(see e.g. Eq.~(7) in Ref.~\cite{MacMinn2016}).
This shows again that the trajectories velocities of the tracers follow  the solid network flux $v_s$, see Eq.~(\ref{eq:traj}).

\section{Discussions and conclusions\label{sec:discussion}}

In the present work, we performed a thorough investigation of the drying kinetics of a charged dispersion in a confined drop.
Our measurements based on Raman micro-spectroscopy lead to measurements of the collective diffusion 
coefficient of the dispersion over a wide concentration range. Fluorescence imaging also reveals a transition from
a liquid dispersion to a solid at a concentration $\varphi_c$ well below the close-packing of the colloids. We also
report that natural convection occurs for such drying experiments, but that mass transport remains dominated by diffusion within 
such confined geometries.

Equations~(\ref{eq:colltrans}) and (\ref{eq:Transport}) in Sec.~\ref{sec:confineddrying} were derived in the general context of binary {\it liquid} dispersions. More specifically, the term $\varphi v$ in Eq.~(\ref{eq:colltrans})
corresponds to the convective flux of colloids, whereas $-D(\varphi) \nabla \varphi$ describes the diffusive part of the flux (in the reference frame of the volume average velocity). 
The collective diffusion coefficient $D(\varphi)$ follows the generalized Stokes-Einstein relation
 \begin{equation}
D(\varphi) = \varphi \frac{k}{\eta_w} \frac{\partial \Pi}{\partial \varphi}, \label{eq:GSE}
\end{equation}
where $\eta_w$ is the water viscosity, $k$ the permeability of the dispersion, and $\Pi(\varphi)$ its osmotic pressure.
This equation shows that the diffusive transport $-D(\varphi) \nabla \varphi$ is actually driven by gradients of osmotic pressure hindered by the hydrodynamic friction of the relative flow solvent/particles~\cite{Russel,Nagele}.  

For a liquid dispersion, the global pressure $P$ often follows a simple mechanical equilibrium $\nabla P = 0$. One 
can thus define the {\it pervadic} pressure $p=P-\Pi$, corresponding to the {\it pore} pressure of the dispersion~\cite{Peppin:05,Peppin:06}. With such a definition, the relative flow solvent/particles can also be written as:
 \begin{equation}
(1-\varphi)(v_f-v_s)  = -\frac{k}{\eta_w} \nabla p\,, \label{eq:darcy}
\end{equation}
thus taking the form of the Darcy equation. This correspondence between Darcy and Fick laws were discussed at length, namely by Peppin {\it et al.} in Refs.~\cite{Peppin:05,Peppin:06} in the context of processes driving suspensions out-of-equilibrium 
(e.g. ultrafiltration, drying, consolidation,\dots). 

For a poro-elastic media, the mechanical equilibrium may not follow  $\nabla P = 0$, and the 
fluid transport should be described by the equations of poro-elasticity. In that case, the relation $\nabla (p + \Pi)=0$ may fail, as mechanical equilibrium of the gel has to be described by a constitutive relation taking also
into account specific boundary conditions. This feature was suggested theoretically recently by Style {\it et al.} in the context of 
unidirectional drying~\cite{Style:11}.
They demonstrated more specifically  that
the colloid transport is still described by Eq.~(\ref{eq:Transport}), but with a collective diffusion coefficient given by
\begin{eqnarray}
D(\varphi) = \varphi \frac{k}{\eta_w}  \frac{3(1-\nu)}{1+\nu} \frac{\partial \Pi}{\partial \varphi},  \label{GSEmod} 
\end{eqnarray}
where $\nu$ is the Poisson ratio of the gel.
This last relation shows how the confinement and the mechanical equilibrium impacts the generalized Stokes Einstein relation~Eq.~(\ref{eq:GSE}) known to describe the relaxation of concentration gradients in a liquid dispersion. However, unidirectional drying corresponds to a highly non-linear configuration 
which should be described using the formalism of large deformation poro-elasticity, see for instance Ref.~\cite{MacMinn2016} for the consolidation of soft porous materials. This complex configuration may prevent from a simple comparison between theory and experiments in order to confirm that such a liquid$\to$solid transition  impacts the generalized Stokes-Einstein relation (from Eq.~(\ref{eq:GSE}) to Eq.~(\ref{GSEmod})).

Our experimental configuration, the drying of a confined gel, leads to similar conclusions obtained by Style {\it et al.}
and more exactly Eq.~(\ref{GSEmod}), but within the framework of 
linear poro-elasticity~\cite{Wang} (see also for instance Ref.~\cite{Chekchaki:13} for the drying of films). Indeed,
we demonstrate in  Appendix~A using linear poro-elasticity modeling, that mass transport is also still described by Eq.~(\ref{eq:Transport}) as above, but with a collective diffusion coefficient given by Eq.~(\ref{GSEmod}) as shown by Style {\it et al.}. The framework of linear poro-elasticity  applies in our geometry as the concentration gradients over the drop are small, see Appendix~A for a demonstration.

We plan in a near future to use again the methodology developed in the present work to measure collective diffusion coefficients over a wider range of concentration (from $\varphi \to 0$ to the consolidation) for colloidal dispersions with a controlled ionic content. 
The combined measurements of the equation of state $\Pi(\varphi)$ and of $D(\varphi)$ may yield to the first direct measurements of the permeability
$k(\varphi)$ using the generalized Stokes-Einstein relation~Eq.~(\ref{eq:GSE}) in the liquid state. The control of the ionic content would also enable us 
to model both $\Pi(\varphi)$ and $D(\varphi)$ knowing the surface charge density of the particles, see e.g. the cell model recently used in the context of ultra-filtration of charged dispersions~\cite{ROA2016}. More importantly, we also plan to perform systematic measurements of the collective diffusion coefficient using scattering techniques for such dispersions. The comparison of these values to the possible measurements using the above methodology would indeed enable us 
to directly test  whether the confinement in such drying experiments impacts or not the generalized Stokes-Einstein relation as suggested above, and demonstrated earlier by Style~{\it et al.}~\cite{Style:11}.

\appendix
    \section{Linear poro-elasticity modeling of the drying of a confined gel} 

The solvent transport for a poro-elastic solid is classically described by the conservation equations Eqs~(\ref{eq:masscons1}-\ref{eq:masscons2}) as above, and the volume average velocity $v$ still obeys $\nabla . v =0$, and thus $v=0$ in our geometry.
The relative flux solvent/colloids is given by the Darcy law Eq.~(\ref{eq:darcy}) with $p$ the pore pressure.
The latter  is related to a constitutive relation between the effective stress of the gel, and its deformations~\cite{Wang}.

Let us consider the drop in the gel phase at $t=t_i$. The mean concentration within the drop is given by $<\varphi> = \varphi_0/\alpha(t_i)$.
The deformations of the elastic network in  this confined geometry, are described by the tensor
\begin{eqnarray}
\mathbf\epsilon = \left| \begin{array}{ccc} 
\partial_r u_r & 0 & 0  \\
0 & u_r/r & 0 \\
0 & 0 & 0 \end{array} \right|,
\end{eqnarray}
where $u_r$ is the radial displacement field. This is consistent with the observations of the trajectories of fluorescent tracers embedded within the gel, see 
Fig.~\ref{fig:Lines}.
We also assume above that the gel recedes freely on the cell surfaces as observed experimentally.

Assuming  small deformations from $t=t_i$ up to time $t$, and for 
small concentration gradients along the drop, one can safely assume that the effective stress $\sigma ' = \sigma + p$ follows a simple linear poro-elastic, isotropic stress-strain relationship:
\begin{eqnarray}
\mathbf\sigma' = \Lambda \xi \mathbf{1} + (\mathcal{M}-\Lambda) \mathbf\epsilon
\end{eqnarray}
with $\mathcal{M}$ and $\Lambda$ the oedometric modulus and Lam\'e's first parameter
(we use here the same notation as in Ref.~\cite{MacMinn2016}).
Mechanical equilibrium $\nabla (\sigma' - p)=0$ writes in this cylindrical geometry 
\begin{eqnarray}
&& \partial_r (r \sigma'_{rr}) - \sigma'_{\theta\theta} = r \partial_r p\,,
\end{eqnarray}
and leads finally to 
\begin{eqnarray}
\mathcal{M} \partial_r \xi = \partial_r p\,.
\end{eqnarray}
We assume above that $\mathcal{M}$, $\Lambda$ are almost homogeneous over the drop and take the values  $\mathcal{M}(<\varphi>)$ and
$\Lambda(<\varphi>)$ during the small deformation. 
These assumptions rely on the fact that  concentration gradients are small over the drop (as observed experimentally), and on the fact that $\mathcal{M}(\varphi)$ and $\Lambda(\varphi)$  do not display abrupt variations with the volume fraction.

With such assumptions, the colloid volume fraction is related to $\xi = \text{tr}(\mathbf\epsilon )$ through
\begin{eqnarray}
\partial_r \varphi(r,t) = - \varphi(r,t)\partial_r \xi\,, 
\end{eqnarray}
and the transport equation Eq.~(\ref{eq:masscons1}) takes finally the following form:
\begin{eqnarray}
\partial_t \varphi = -\frac{1}{r}\partial_r \left(r \frac{k(\varphi) \mathcal{M}(\varphi)}{\eta_w} \partial_r \varphi\right)\,. 
\end{eqnarray}
Boundary condition at $r=R$ still  follows Eq.~(\ref{BC1}), and the collective diffusion coefficient 
within the gel is thus:
\begin{eqnarray}
D(\varphi) &&= \frac{k(\varphi) \mathcal{M}(\varphi)}{\eta_w}\,, 
\end{eqnarray}
also known as the consolidation coefficient~\cite{Wang}.
The oedometric modulus $\mathcal{M}$ is related to the osmotic compressibility (as measured classically using osmostic shock techniques for instance), and the above relation finally writes as Eq.~(\ref{GSEmod}), see also Ref.~\cite{Style:11} for more details.



\begin{acknowledgments}
We thank G. Ovarlez, L. Pinaud, J. Leng, F. Doumenc, B.~Guerrier, Y. Hallez, and D. Ou-Yang for useful discussions. 
We also thank Solvay and CNRS for fundings. The research leading to these results received also fundings from {\it Agence Nationale de la Recherche} for the grant {\it EVAPEC} (ANR-13-BS09-0010).
This work was supported by the LabEx AMADEus (ANR-10LABX-42) in the framework of IdEx Bordeaux (ANR-10-IDEX03-02), i.e., the Investissements d'Avenir program of the French government managed by the {\it Agence Nationale de la Recherche}.
\end{acknowledgments}


%

\end{document}